\documentclass[9pt,conference]{IEEEtran}
\IEEEoverridecommandlockouts

\usepackage{cite}
\usepackage{amsmath,amssymb,amsfonts}
\usepackage{algorithmic}
\usepackage{graphicx}
\usepackage{textcomp}
\usepackage{xcolor}
\usepackage{array}
\usepackage{url}
\usepackage{multirow}
\usepackage{balance}
\usepackage{bbm}
\usepackage{caption}
\usepackage{subfigure}
\usepackage{setspace}
\usepackage[T1]{fontenc}
\usepackage{threeparttable}
\usepackage{booktabs}
\usepackage{makecell}

\usepackage{tablefootnote}

\def\BibTeX{{\rm B\kern-.05em{\sc i\kern-.025em b}\kern-.08em
    T\kern-.1667em\lower.7ex\hbox{E}\kern-.125emX}}
\begin{document}

\title{Unified Audio Event Detection}


\author{\IEEEauthorblockN{
Yidi Jiang$^{1,3}$, Ruijie Tao$^{1 \dag}$\thanks{$\dag$ Corresponding Author}, Wen Huang$^{2}$, Qian Chen$^{3}$, Wen Wang$^{3}$}
\IEEEauthorblockA{\textit{$^1$National University of Singapore, Singapore~~~ $^{2}$Shanghai Jiao Tong University, Shanghai, China~~~ $^{3}$Speech Lab, Alibaba Group}}
\IEEEauthorblockA{
\{yidi\_jiang, ruijie.tao\}@u.nus.edu~~~ holvan@sjtu.edu.cn~~~ \{tanqing.cq, w.wang\}@alibaba-inc.com
}
}


\maketitle

\begin{abstract}
Sound Event Detection (SED) detects regions of sound events, while Speaker Diarization (SD) segments speech conversations attributed to individual speakers. In SED, all speaker segments are classified as a single speech event, while in SD, non-speech sounds are treated merely as background noise. Thus, both tasks provide only partial analysis in complex audio scenarios involving both speech conversation and non-speech sounds. 
In this paper, we introduce a novel task called Unified Audio Event Detection~(UAED) for comprehensive audio analysis. UAED explores the synergy between SED and SD tasks, simultaneously detecting non-speech sound events and fine-grained speech events based on speaker identities. To tackle this task, we propose a Transformer-based UAED~(T-UAED) framework and construct the UAED Data derived from the Librispeech dataset and DESED soundbank. Experiments demonstrate that the proposed framework effectively exploits task interactions and substantially outperforms the baseline that simply combines the outputs of SED and SD models. T-UAED also shows its versatility by performing comparably to specialized models for individual SED and SD tasks on DESED and CALLHOME datasets.


\end{abstract}


\section{Introduction}
In the area of audio signal analysis, Sound Event Detection~(SED)~\cite{Turpault2019_DCASE,mesaros2021sound,mesaros2016metrics} aims to detect the presence of sound events, while Speaker Diarization (SD) seeks to demarcate the speech segments of each speaker to answer the ``who spoke when'' question in a multi-speaker scenario~\cite{chen2023attention,cheng2023target,medennikov2020target}. 
Both SED and SD can be conceptualized as tasks where audio signals are analyzed to detect and label specific types of events or segments at the frame level.


However, in complex audio environments involving both speech conversation and non-speech sounds, e.g. on the train or in a factory,  current SED and SD approaches provide only partial audio analysis. 
More specifically, SED focuses on general sounds and often classifies speech segments from different speakers as a single speech event. Conversely, SD focuses solely on speech event and disregards non-speech sounds, merely treating non-speech sounds as background noise without explicit modeling them. 



\begin{figure}[!t]
    \centering
      \includegraphics[scale=0.62]{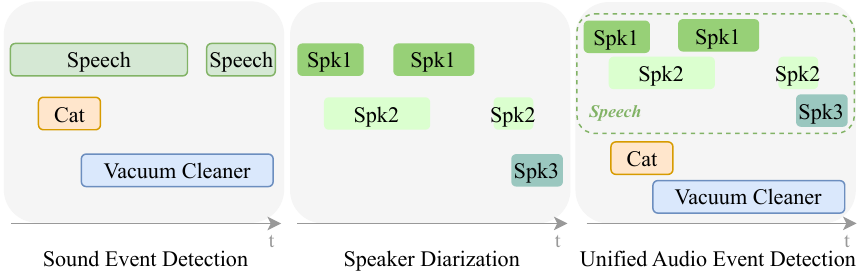}
    \caption{{Illustration of the differences between our proposed UAED task and the traditional SED and SD tasks. 
    UAED aims to simultaneously identify the regions of both non-speech sound events and speech segments from individual speakers, thereby providing a holistic view of the soundscape.}}
    \label{fig:illustration}
\end{figure}

We consider that SED and SD have synergistic interactions and hypothesize that in complex audio environments, accurately modeling non-speech sound events instead of merely treating them as background noise can improve the detection of speaker-related events. 
On the other hand, accurately detecting individual speech segments for each speaker should improve overall speech event detection in SED under long speech conversation scenarios.
Motivated by this hypothesis, we propose a new task termed \textbf{Unified Audio Event Detection~(UAED)} to  provide unified and comprehensive understanding and analysis of audio data.
As illustrated in Figure~\ref{fig:illustration}, 
UAED can be considered as simultaneously conducting SED and SD on the input audio, in which segments of each non-speech sound event are detected and speech event is divided into fine-grained segments based on speaker identities. The output of UAED is the start and end time of each audio event, including non-speech sound events and speech segments of individual speakers. 

In this paper, we propose a Transformer-based unified Audio Event Detection (T-UAED) framework as a closed-set, supervised approach to the UAED task. T-UAED employs a combination of a sound encoder and a speaker encoder as the front-end auditory encoders to extract both general sound and speaker-specific acoustic information. To achieve the UAED target, T-UAED uses \textit{UAED Queries} and the Transformer encoder-decoder architecture. Each \textit{UAED Query} corresponds to each identifiable audio event to furnish information for all the audio events in the closed set. 
The coordination of UAED Queries and Transformer encoder-decoder enables capturing complex temporal patterns in the audio stream, by leveraging the parallel nature of the cross-attention structure.

Our contribution can be summarized as follows:
\begin{enumerate}
    \item We introduce a novel Unified Audio Event Detection~(UAED) task for comprehensive audio analysis. We propose a Transformer-based Unified Audio Event Detection~(T-UAED) framework to detect the regions of both non-speech sound events and speaker-aware speech events.
    \item We devise a data simulation strategy and evaluation metrics tailored for the UAED task to establish the benchmark. Experiments on this benchmark demonstrate the effectiveness of the T-UAED framework for the UAED task.
    \item The T-UAED framework also shows performance comparable to specialized models dedicated to SED and SD tasks, proving its adequacy to support these two tasks. 
\end{enumerate}


\section{Transformer-based Unified Audio Event Detection~(T-UAED)}
We propose a Transformer-based Unified Audio Event Detection~(T-UAED) framework to detect both non-speech sound events and speaker-aware speech events within the input audio. The overview architecture of T-UAED is depicted in Figure~\ref{fig:overview}, which consists of auditory encoders with adapters, \textit{UAED Queries}, and a Transformer encoder-decoder architecture. The unified audio event detection outputs are binary sequences $Y \in \{0, 1\}^{N \times T}$, where $1$ represents the audio event being active and $0$ represents the audio event being absent. $N$ is the number of audio events comprising both non-speech sound events and speaker-aware events, and $T$ is the number of frames of the input audio. 

The auditory encoders first map the input audio to a feature sequence $F_U \in \mathbb{R}^{T \times D}$, where $D$ is the feature dimension. Then, the Transformer encoder-decoder takes $F_U$ and UAED Queries as inputs and outputs the prediction for each audio event. 
Specifically, the Transformer encoder takes $F_U$ as input and outputs the frame-level speech representation $F_{enc} \in \mathbb{R}^{T \times D}$. The Transformer decoder takes \textit{UAED Queries} $Q=[q_1,q_2,\ldots,q_N] \in \mathbb{R}^{N \times D}$ as query and $F_{enc}$ as key and value in cross attention, and outputs $F_{dec} \in \mathbb{R}^{N \times D}$. Finally, we perform a dot product operation between the decoder output $F_{dec}$ and the encoder output $F_{enc}$ and apply a sigmoid operation to obtain the prediction sequence $\hat{\mathrm{Y}} \in (0, 1)^{N \times T}$. The value of $\hat{\mathrm{Y}}$ denotes the probability of audio event occurrence at each frame. 

\subsection{Auditory encoders with adapters}
\begin{figure}[!t]
    \centering
      \includegraphics[scale=0.55]{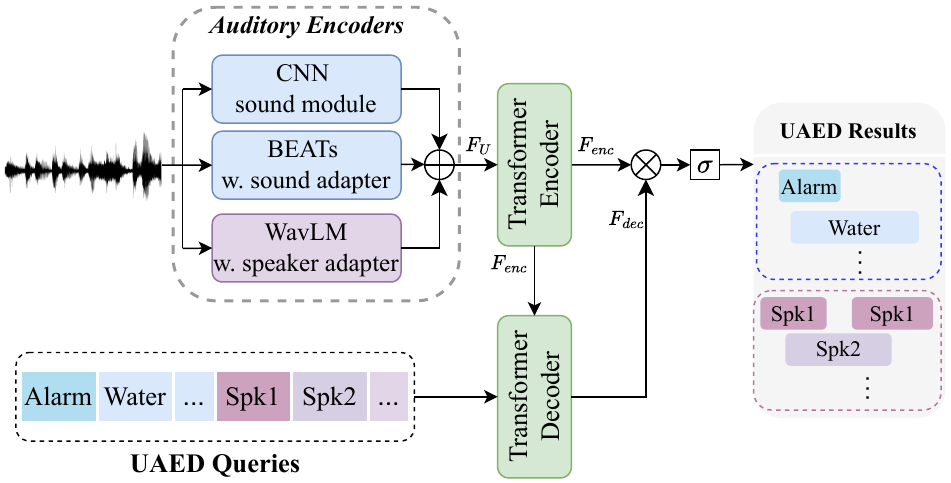}
    \caption{The overview of our Transformer-based unified Audio Event Detection~(T-UAED) framework. It outputs the occurrence regions for each audio event comprising both non-speech sound events and speaker-aware speech events. In this framework, $\oplus$ symbolizes feature fusion layer.
    $\otimes$ and $\sigma$ represent the dot product and sigmoid operation, respectively. }
    \label{fig:overview}
\end{figure}

In order to extract both general sound and speaker-specific acoustic information from the input audio, we employ a combination of sound encoder and speaker encoder. As illustrated in the block of Auditory Encoders in Figure~\ref{fig:overview}, we follow the design of CRNN-BEATs~\cite{Turpault2019_DCASE} and use a pretrained BEATs encoder~\cite{chen2022beats} with a sound adapter together with a paralleled CNN sound module, as the sound encoder. We use a pretrained WavLM encoder~\cite{chen2022wavlm} with a speaker adapter as the speaker encoder.

Within the sound encoder, BEATs~\cite{chen2022beats} is trained to capture high-level semantic information of sound through iterative self-supervised learning. The self-supervised learning process consists of audio tokenization, masking audio tokens, and predicting masked tokens.
The CNN sound module captures detailed local spectral and temporal variations in the audio signal.
As the speaker encoder, the WavLM encoder~\cite{chen2022wavlm} is designed to learn unified speech representations from vast amounts of unlabeled speech data, ensuring robustness of the frame-level speaker-related representation.

The auditory features extracted by these three encoders are complementary, catering well to representing general audio inputs that contain varied sound and speaker information. Given that both BEATs and WavLM encoders output features at a frame rate of 50Hz, we align the CNN features to this temporal resolution through pooling. The features from all three modules are then concatenated frame-by-frame along the feature dimension. This concatenated output is subsequently processed by a 1-D convolutional layer to downsample both the features and the frame rate. 

Inspired by the design of the semantic adapter and the acoustic adapter in~\cite{hu2024wavllm}, we employ a sound adapter and a speaker adapter to the pretrained BEATs and WavLM encoders for parameter-efficient fine-tuning. Both adapters consist of three components: two 1-D convolutional layers for downsampling and temporal alignment of the encoder outputs, a down-up bottleneck adapter~\cite{houlsby2019parameter}, and a linear projection layer.
The sound adapter processes the output from the BEATs encoder, while the speaker adapter takes a weighted sum of the hidden states from all layers of WavLM, where the weights are learnable.

\subsection{UAED Queries and Transformer encoder-decoder}
In order to simultaneously detect sound events and speaker-aware events, we introduce \textit{UAED Queries} to furnish information for the $N$ audio events within a closed-set, supervised approach for UAED. Each query corresponds to a specific audio event with the dimension $D$. Learnable embeddings are assigned specifically for each non-speech sound event. Additionally, for each speaker-aware event, we utilize the speaker embeddings extracted from the pre-enrolled speech. In this study, we adopt the high quality pre-trained ECAPA-TDNN encoder~\cite{desplanques2020ecapa} to extract speaker embeddings as the \textit{UAED Queries} corresponding to speaker-aware events.

Our model is based on the Transformer encoder-decoder architecture as introduced in~\cite{vaswani2017attention}. This architecture benefits from the self-attention and cross-attention mechanisms, which are adept at capturing complex temporal patterns in audio data and aligning them effectively with \textit{UAED Queries}. The integration of \textit{UAED Queries} and queried Transformer enables our model to accurately identify and detect active regions of each audio event, providing a robust and feasible solution for the UAED task.

\subsection{Loss function}
The learning targets of our framework are frame-wise binary ground truth labels $Y \in \{0,1\}^{N \times T}$ for the $N$ audio events.
For each audio event, we utilize binary cross-entropy loss to train our model, as defined in Equation~\ref{eq:loss}. $\hat{y}_{n,t}$ and $y_{n,t}$ represent the predicted and ground truth labels of the $n^{th}$ event for the $t^{th}$ audio frame, where $n \in [1, N]$ and $t \in [1, T]$. The loss function is to minimize the difference between predicted and ground truth labels.

\begin{equation}
    \mathcal{L} = -\frac{1}{N*T} \sum_{n=1}^{N} \sum_{t=1}^{T} (y_{n,t} \cdot log \hat{y}_{n,t} + (1-y_{n,t}) \cdot log (1- \hat{y}_{n,t}))
    \label{eq:loss}
\end{equation}

\begin{table*}[!t]
    \centering
   \caption{The overall analysis for our proposed T-UAED framework. All the involved models are trained and evaluated on simulated \textit{UAED data}. Models A-D are tailored specifically for either SED or SD task. \textbf{Models A+B}, \textbf{E-H} and our proposed \textbf{UAED} can achieve UAED task. We also report the SED and SD results derived from the UAED outputs.
   L.E. represents learnable embedding for each sound event as UAED Queries. Spk Emb. represents speaker embedding for each speaker-aware event as UAED Queries}
    \begin{tabular}{c|cc|cc|c|cc}
      \toprule
      \multirow{2}{*}{\textbf{ID}} & \multicolumn{2}{c|}{\textbf{Model}} & \multicolumn{2}{c|}{\textbf{SED Results}} & \textbf{SD Results} & \multicolumn{2}{c}{\textbf{UAED Results}} \\
      \cmidrule(lr){2-3} \cmidrule(lr){4-5} \cmidrule(lr){6-6} \cmidrule(lr){7-8}
      & Auditory Encoder & UAED Queries & IB-F1~($\uparrow$) & SB-F1~($\uparrow$) & DER~(MS/FA/SC)~($\downarrow$) & IB-F1~($\uparrow$) & SB-F1~($\uparrow$) \\
      \midrule
      A & CNN+BEATs & L.E. & 0.652 & 0.828 & - & - & - \\
      B & CNN+BEATs+WavLM & L.E. & 0.628 & 0.806 & - & - & - \\
      C & WavLM & Spk Emb. & - & - & 10.07~(4.12/3.63/2.32) & - & - \\
      D & CNN+BEATs+WavLM & Spk Emb. & - & - & 11.47~(5.46/3.98/2.03) & - & - \\
      \midrule 
      A+B & (CNN+BEATs)+(WavLM) & (L.E.)+(Spk Emb.) & 0.647 & 0.828 & 10.07~(4.12/3.63/2.32) & 0.608 & 0.796 \\
      \textbf{T-UAED} & \textbf{CNN+BEATs+WavLM} & \textbf{L.E.+Spk Emb.} & \textbf{0.658} & \textbf{0.844} & \textbf{7.24~(2.19/1.61/3.43)} & \textbf{0.652} & \textbf{0.824} \\
      \midrule
      E & w/o CNN & L.E.+Spk Emb. & 0.605 & 0.827 & 8.45~(3.45/2.35/2.65) & 0.603 & 0.809 \\
      F & w/o BEATs & L.E.+Spk Emb. & 0.651 & 0.824 & 9.36~(3.37/2.67/3.31) & 0.629 & 0.797 \\ 
      G & w/o sound adapter & L.E.+Spk Emb. & 0.644 & 0.831 & 8.95~(3.32/2.19/3.43) & 0.631 & 0.81 \\
      H & w/o speaker adapter & L.E.+Spk Emb. & 0.647 & 0.834 & 9.81~(3.98/2.12/3.71) & 0.626 & 0.804 \\
      
      \bottomrule
    \end{tabular}
    \label{tab:uad}
\end{table*}

\section{Experimental Setting}
In this section, we detail the datasets, evaluation metrics, and experimental setup for evaluating the proposed T-UAED framework.

\subsection{Dataset simulation}
\label{subsec:dataset}
To address the scarcity of real-world data for our proposed UAED task, we develop an online data simulation strategy tailored for this task and construct \textit{UAED Data}. \textit{UAED Data} consists of mixtures of speech conversation utterances and various sound events with labels for supervised learning.
Since the SED task typically operates in a closed-set scenario, this study also focuses on a closed-set scenario where the categories of sound events and the fixed number of speakers per utterance are known.

First, we follow the recipe~\footnote{https://github.com/BUTSpeechFIT/EEND\_dataprep/} proposed in~\cite{landini2023multi} to simulate 3-speaker speech conversation datasets from Librispeech~\cite{panayotov2015librispeech}. To create datasets that closely resemble real-world speech conversations, we utilize conversation statistics computed on the DIHARD II development set~\cite{ryant2019second} to synthesize a speech conversation dataset with 3 speakers, among which 500 hours data are used for training and 5 hours are used as the evaluation set. 

Next, for each 32-second speech utterance, we randomly select 0 or 1 background event and 1 or 2 foreground events from the DESED soundbank~\cite{Turpault2019_DCASE}. The soundbank~\footnote{https://zenodo.org/records/3702397} contains 9 classes of isolated non-speech sound events.
The background sound events are randomly truncated to 10-30 seconds duration from the five long sound event classes~(water/cleaner/frying/blender/toothbrush). The foreground events are randomly truncated to 1-5 seconds duration
from 9 non-speech sound events~(alarm ringing/cat/dish/dog and etc)
Each event audio is integrated into the speech utterance at a random start time, while varying the signal-to-noise ratio~(SNR) randomly from 10 dB to 15 dB for the background events and varying SNR randomly from 0 dB to 5 dB for the foreground events.


\subsection{Evaluation metric}
We adopt intersection-based F1 scores~(IB-F1)~\cite{bilen2020framework} and segment-based F1 scores~(SB-F1)~\cite{mesaros2016metrics} as metrics for the UAED task. Each metric targets measuring distinct aspects of the system's performance. The metrics are computed using the sed\_scores\_eval library\footnote{https://github.com/fgnt/sed\_scores\_eval/}. 
\cite{bilen2020framework} present the intersection criterion approach, which refines the definitions of true positives (TPs) and false positives (FPs), as well as cross-triggers (CTs) in polyphonic sound event detection (SED). This metric quantifies the overlap between ground truth and system prediction, supporting both merged and interrupted detections. It ensures that evaluations are unaffected by the time scale of event durations, hence provides the necessary flexibility to accommodate subjective labeling and to mitigate biases associated with event durations.
Conversely, segment-based metrics~\cite{mesaros2016metrics} assess the system's output against the reference within brief time intervals. The length of these intervals is adjustable, allowing for tailored resolution to suit specific application needs.

\subsection{Implementation details}
The T-UAED framework is implemented with PyTorch and optimized with the Adam optimizer. The initial learning rate is set to $10^{-4}$ and decreased by 5\% for each epoch. The dimension $D$ of audio feature $F_U$ and UAED Queries are both set to 192. Both Transformer encoder and decoder consist of 6 layers with 8 attention heads.

\section{Results and Analysis}
\label{sec:res}
\subsection{T-UAED evaluation}

\subsubsection{Baselines}
\textbf{Model A, B} as baselines are dedicated to sound event detection. 
Model A utilizes a CNN sound module and a BEATs sound encoder as the auditory encoder, while Model B employs the same auditory encoder as T-UAED.
The UAED Queries in Model A and Model B are learnable embeddings specific to each sound event. 
Serving as the sound event branch of the T-UAED framework, Model A and B solely outputs sound event regions, categorizing segments from different speakers as speech event.
\textbf{Model C, D} as other baselines are designed exclusively for the speaker diarization task. 
Model C utilizes WavLM speaker encoder while Model D employs the same auditory encoder as T-UAED.
Both of them operate as the speaker branch of the T-UAED framework, with speaker embeddings serving as the UAED queries. 
Model C and Model D output only the speech regions for each speaker, treating other non-speech sound events as noise. 
\textbf{Model A+B} serves as a UAED baseline. This model employs two distinct models and integrates the non-speech sound event results from Model A and the speaker-aware event results from Model B to produce the final UAED results. 


\subsubsection{Analysis of UAED results}
All the models in Tabel~\ref{tab:uad} are trained on \textit{UAED data} (Section~\ref{subsec:dataset}).
In Table~\ref{tab:uad}, the first group comprises single-task models that solely perform either SED or SD task. The second and third groups of Table~\ref{tab:uad} are dedicated to UAED models, which not only output comprehensive UAED results but also provide separated SED and SD results derived from the UAED results. The SED results are calculated by aggregating all speaker-aware events from the UAED results, treated as speech events, along with other non-speech sound events. Conversely, the SD results are derived from the speaker-aware events to calculate the Diarization Error Rate (DER), which comprises the summation of Miss~(MS), False Alarm~(FA), and Speaker Confusion~(SC)~\cite{friedland2011icsi}.

The comparative analysis of \textbf{Model A vs. Model B} and \textbf{Model C vs. Model D} highlights that WavLM features do not contribute to sound event detection, and BEATs features do not suit speaker-related events. The UAED results for \textbf{Model A+B} are derived by combining non-speech sound event results from \textbf{Model A} with speaker-aware speech events from \textbf{Model B}. Using the same front-end as Model A+B, our proposed \textbf{T-UAED} system yields superior performance compared to \textbf{Model A+B} for the UAED task and the specific SED and SD tasks. This observation underscores the synergistic interdependence between sound event detection and speaker diarization and demonstrates the advantages of jointly modeling SED and SD for the UAED task. Accurately modeling background noise as sound events can significantly enhance the detection of speaker-related events. Integrating the two tasks in UAED facilitates a more comprehensive understanding and processing of audio data and improves overall system effectiveness in diverse audio environments.

\textbf{Models E-H} serve as ablation studies to evaluate the contribution of each component, including the CNN sound module, BEATs sound encoder, sound adapter and speaker adapter, to overall system performance.
Ablation results show that omitting the CNN sound module substantially degrades sound event detection, indicating that the CNN sound module plays a more important role to this task benefiting from its finer temporal resolution compared to the patch-wise BEATs features.
As expected, removing the sound adapter from the BEATs encoder and excluding the speaker adapter from the WavLM speaker encoder negatively impacts sound event detection (SED results) and speaker-aware events detection (SD results), respectively.

\subsection{T-UAED-SED evaluation}
To demonstrate the effectiveness of the SED branch within the T-UAED framework (denoted \textbf{T-UAED-SED}), we compare T-UAED-SED with SOTA SED systems on the widely adopted DESED dataset~\cite{Turpault2019_DCASE} in DCASE challenge. We evaluate performance using two polyphonic sound detection score (PSDS) metrics, PSDS1 and PSDS2, following recent works~\cite{ebbers2022threshold}.

For training, we adopt the same semi-supervised framework and feature front-end as the CRNN-BEATs baseline~\cite{Turpault2019_DCASE} and only modify the context backend. 
Different from the RNN architecture used in the backend of CRNN-BEATs, our approach instead incorporates learnable embeddings as UAED Queries and a Transformer architecture. 
This modification yields superior performance, achieving PSDS1 score of 0.517 and PSDS2 score of 0.785, as shown in Table~\ref{tab:sed}. 
In comparison, ATST-SED~\cite{shao2024fine}, which finetunes a large pretrained ATST-frame encoder with extra AudioSet-2M dataset~\cite{gemmeke2017audio} and adopts post-processing strategies, achieves SOTA performance.
Compared to other methods without extra encoder finetuning in Table~\ref{tab:sed}, T-UEAD-SED delivers comparable performance, underscoring the effectiveness and adaptability of our approach for SED task.

\begin{table}[h]
    \centering
   \caption{PSDS1 and PSDS2 results compared with other SED methods trained on DESED dataset. $^{*}$ denotes the pretrained ATST-frame encoder is further finetuned on the extra AudioSet-2M data.}
    \begin{tabular}{c|cc}
      \toprule
       \textbf{Model} & \textbf{PSDS1}~($\uparrow$) & \textbf{PSDS2}~($\uparrow$) \\
        \midrule 
        AST-SED~\cite{li2023ast} & 0.514 & - \\
        ATST-SED~(BEATs freeze)~\cite{shao2024fine} & 0.501 & 0.755 \\
        ATST-SED$^{*}$~(ATST finetune)~\cite{shao2024fine} & 0.583 & 0.810 \\
        FDY-LKA-CRNN~\cite{kim2023label} & 0.526 & 0.782 \\
        CRNN-BEATs~\tablefootnote{https://github.com/DCASE-REPO/DESED\_task/tree/master/recipes/\\dcase2023\_task4\_baseline} & 0.500 & 0.762 \\
        \textbf{T-UAED-SED} & 0.517 & 0.785 \\
      \bottomrule
    \end{tabular}
    \label{tab:sed}
\end{table}

\subsection{T-UAED-SD evaluation}

To demonstrate the effectiveness of the SD branch within the T-UAED framework (denoted by \textbf{T-UAED-SD}), we conduct comparison experiments with state-of-the-art speaker diarization systems. The traditional two-stage speaker diarization training process includes a pre-training stage with the simulated dataset and an adaptation stage with the real-world dataset.
To ensure a fair comparison, we follow the simulation strategy in~\cite{czy-taslp} to generate two subsets of 2 speakers and 3 speakers, using CALLHOME dataset Part1 statistics. 

In Table~\ref{tab:sd}, we compare the DER performance on the CALLHOME Part 2 2-spk and 3-spk subsets with SOTA and competitive SD systems. When speaker embeddings are extracted from \textit{oracle} pre-enrolled speech for each speaker, our \textbf{T-UAED-SD~(oracle)} achieves impressive DER of 6.80\% and 8.51\% for 2 and 3 speakers, respectively. While adopting a clustering-based system~\cite{wang2023wespeaker} to extract speaker embeddings from the estimated reference speech for each speaker, our \textbf{T-UAED-SD~(cluster)} achieves 8.17\% and 11.76\% DER for 2 and 3 speakers, respectively, and maintains comparable performance with SOTA SD systems. 

\begin{table}[!ht]
\caption{DER~(\%) results comparison on the CALLHOME Part 2 2-spk and 3-spk subsets. Lower is better. \textbf{T-UAED-SD~(oracle)} denotes T-UAED-SD with the ground-truth enrolled speech as speaker embedding for speaker diarization, \textbf{T-UAED-SD~(cluster)} denotes T-UAED-SD with speaker embeddings extracted from the estimated reference speech by clustering front-end.}
\label{tab:sd}
\centering
\begin{threeparttable}
\begin{resizebox}{0.95\columnwidth}{!} {
\begin{tabular}{c|cc}
\toprule
\textbf{Method} & \textbf{\makecell{CALLHOME \\ 2spk}} & \textbf{\makecell{CALLHOME \\ 3spk}} \\

\midrule
x-vector clust.~\cite{horiguchi2020end} & 11.53 & 19.01 \\
clut. frontend~\cite{wang2023wespeaker} &15.60& 21.25 \\
BLSTM-EEND~\cite{fujita2019end} & 26.03 & -\\
SA-EEND~\cite{horiguchi2020end,fujita2019end} & 9.54 & 14.00 \\
SC-EEND~\cite{fujita2020neural} & 8.86 & - \\
EEND-EDA~\cite{horiguchi2020end} & 8.07 & 13.92 \\
EEND-EDA $^\dagger$ & 8.32 & 17.07 \\
TS-VAD~\cite{medennikov2020target} & 9.51 & -\\
MTFAD~\cite{medennikov2020target} & 7.82 & - \\
AED-EEND~\cite{czy-taslp} & 7.75 & 12.87 \\
\midrule
\textbf{T-UAED-SD~(oracle)} & \textbf{6.80} & \textbf{8.51} \\
\textbf{T-UAED-SD~(cluster)} & 8.17 & \textbf{11.76} \\
\bottomrule
\end{tabular}
} \end{resizebox}
\begin{tablenotes}\footnotesize
\item $^\dagger$: our implementation.
\end{tablenotes}
\end{threeparttable}
\end{table}


\section{Conclusion}
In this paper, we introduce a novel task termed Unified Audio Event Detection (UAED), designed to integrate and explore the interactions between sound event detection and speaker diarization tasks. The UAED task aims to detect both non-speech sound events and speaker-aware speech events, providing a comprehensive analysis in various audio environments. Additionally, we have developed the Transformer-based unified Audio Event Detection (T-UAED) framework as a benchmark to effectively address the UAED task.
Our findings confirm that accurately modeling background noise as sound events significantly enhances the detection of speaker-related speech events.
This work establishes a benchmark for the UAED task but is currently constrained by its closed-set setup. Future studies will focus on improving the framework's generalization to handle unseen sound events and unknown speakers.

\clearpage

\bibliographystyle{IEEEtran}
\bibliography{refs}

\end{document}